\documentclass{amsarta}
\usepackage[dvips]{epsfig}
\usepackage{graphicx}
\usepackage{amscd}
\usepackage{amsmath}
\usepackage{amsxtra}
\usepackage{amsfonts}
\usepackage{amssymb}
\usepackage{latexsym}

\theoremstyle{definition}

\theoremstyle{remark}

\renewcommand{\theclaim}{\textup{\theclaim}}

\newtheorem*{acknowledgements}{Acknowledgements}

\renewcommand{\theenumi}{\roman{enumi}}
\renewcommand{\labelenumi}{(\theenumi)}
\numberwithin{equation}{section}

\def\openone

{\mathchoice

{\hbox{\upshape \small1\kern-3.3pt\normalsize1}}

{\hbox{\upshape \small1\kern-3.3pt\normalsize1}}

{\hbox{\upshape \tiny1\kern-2.3pt\SMALL1}}

{\hbox{\upshape \Tiny1\kern-2pt\tiny1}}}

\makeatletter

\newbox\ipbox

\newcommand{\diracb}[1]{\left\langle #1\mathrel{\mathchoice

{\setbox\ipbox=\hbox{$\displaystyle \left\langle\mathstrut
#1\right.$}

\vrule height\ht\ipbox width0.25pt depth\dp\ipbox}

{\setbox\ipbox=\hbox{$\textstyle \left\langle\mathstrut
#1\right.$}

\vrule height\ht\ipbox width0.25pt depth\dp\ipbox}

{\setbox\ipbox=\hbox{$\scriptstyle \left\langle\mathstrut
#1\right.$}

\vrule height\ht\ipbox width0.25pt depth\dp\ipbox}

{\setbox\ipbox=\hbox{$\scriptscriptstyle \left\langle\mathstrut
#1\right.$}

\vrule height\ht\ipbox width0.25pt depth\dp\ipbox}

}\right. }

\newcommand{\dirack}[1]{\left. \mathrel{\mathchoice

{\setbox\ipbox=\hbox{$\displaystyle \left.\mathstrut
#1\right\rangle$}

\vrule height\ht\ipbox width0.25pt depth\dp\ipbox}

{\setbox\ipbox=\hbox{$\textstyle \left.\mathstrut
#1\right\rangle$}

\vrule height\ht\ipbox width0.25pt depth\dp\ipbox}

{\setbox\ipbox=\hbox{$\scriptstyle \left.\mathstrut
#1\right\rangle$}

\vrule height\ht\ipbox width0.25pt depth\dp\ipbox}

{\setbox\ipbox=\hbox{$\scriptscriptstyle \left.\mathstrut
#1\right\rangle$}

\vrule height\ht\ipbox width0.25pt depth\dp\ipbox}

} #1\right\rangle}

\newcommand{\bz}{\mathbb{Z}}
\newcommand{\br}{\mathbb{R}}

\usepackage{graphicx}

\def\blfootnote{\xdef\@thefnmark{}\@footnotetext}

\begin{document}
\title[Comparison of Discrete and Continuous Wavelet Transforms]{Comparison of Discrete and Continuous Wavelet Transforms}
\author{Palle E.T. Jorgensen}
\address[Palle E.T. Jorgensen]{Department of Mathematics\\
The University of Iowa\\
14 MacLean Hall\\
Iowa City, IA 52242}
\email{jorgen@math.uiowa.edu}

\author{Myung-Sin Song}
\address[Myung-Sin Song]{Department of Mathematics and Statistics\\
Southern Illinois University Edwardsville\\
Box 1653, Science Building\\
Edwardsville, IL 62026}
\email{msong@siue.edu}\ 

\thanks{Work supported in part by the U.S. National Science
Foundation. The full version with figures can be found at 
http://www.siue.edu/~msong/Research/ency.pdf}

\maketitle 
\tableofcontents

\section*{Glossary}
\label{sec2:Glo}
This \emph{glossary} consists of a list of terms used inside the paper: In 
mathematics, in probability, in engineering, and on occasion in
physics. To clarify the seemingly confusing use of up to four different
names for the same idea or concept, we have further added informal
explanations spelling out the reasons behind the differences in current
terminology from neighboring fields.

\bigskip
 
\noindent
\textsc{Disclaimer:} 
     This glossary has the structure of four columns. A number of terms are
listed line by line, and each line is followed by explanation. Some ``terms''
have up to four separate (yet commonly accepted) names. 
\bigskip

\begin{minipage}[t]{0.24\textwidth}{\scshape mathematics}\end{minipage}\kern0.01\textwidth\begin{minipage}[t]{0.24\textwidth}{\scshape probability}\end{minipage}\kern0.01\textwidth\begin{minipage}[t]{0.24\textwidth}{\scshape engineering}\end{minipage}\kern0.01\textwidth\begin{minipage}[t]{0.24\textwidth}{\scshape physics}\end{minipage}\smallskip

\hrule\medskip

\noindent
\begin{minipage}[t]{0.24\textwidth}{\bfseries\raggedright function\\(measurable)}\end{minipage}\kern0.01\textwidth\begin{minipage}[t]{0.3\textwidth}{\bfseries\raggedright random variable}
\end{minipage}\kern0.01\textwidth\begin{minipage}[t]{0.24\textwidth}{\bfseries\raggedright signal}\end{minipage}\kern0.01\textwidth\begin{minipage}[t]{0.24\textwidth}{\bfseries\raggedright state}
\end{minipage}

\begin{quote} 
       Mathematically, functions may map between any two sets, say, from 
$X$ to $Y$; but if $X$ is a probability space (typically called $\Omega$), 
\end{quote}

\begin{minipage}[t]{0.24\textwidth}{\scshape mathematics}\end{minipage}\kern0.01\textwidth\begin{minipage}[t]{0.24\textwidth}{\scshape probability}\end{minipage}\kern0.01\textwidth\begin{minipage}[t]{0.24\textwidth}{\scshape engineering}\end{minipage}\kern0.01\textwidth\begin{minipage}[t]{0.24\textwidth}{\scshape physics}\end{minipage}\smallskip

\hrule\medskip

\begin{quote} 
it comes
with a $\sigma$-algebra
 $\mathcal{B}$ of measurable sets, and probability measure
$P$. Elements $E$ in $\mathcal{B}$ are called events, and P(E) 
\end{quote}

\begin{quote} 
\noindent
the probability of
$E$. Corresponding measurable functions with values 
in a vector space are
called random variables,
a terminology 
which suggests a stochastic
viewpoint. The function values of a 
random variable
may represent the outcomes
of an experiment, for example ``throwing of a die.'' 

\quad
       Yet, function theory is widely used in engineering where
functions are typically thought of as signal. In this case, $X$ may be the
real line for time, or $\mathbb{R}^d$. Engineers visualize functions
as signals. A particular signal may have a stochastic component, and this
feature simply introduces an extra stochastic variable into the ``signal,''
for example noise. 

\quad
       Turning to physics, in our present application, the physical
functions will be typically be in some $L^2$-space, and $L^2$-functions with 
unit norm represent quantum mechanical ``states.''
\end{quote}

\bigskip

\noindent
\begin{minipage}[t]{0.24\textwidth}{\bfseries\raggedright sequence (incl.\\vector-valued)}\end{minipage}\kern0.01\textwidth\begin{minipage}[t]{0.24\textwidth}{\bfseries\raggedright random walk}
\end{minipage}\kern0.01\textwidth\begin{minipage}[t]{0.24\textwidth}{\bfseries\raggedright time-series}\end{minipage}\kern0.01\textwidth\begin{minipage}[t]{0.24\textwidth}{\bfseries\raggedright measurement}\end{minipage}

\begin{quote}
         Mathematically, a sequence is a function defined on the integers $\mathbb{Z}$
or on subsets of $\mathbb{Z}$, for example the natural numbers 
$\mathbb{N}$. Hence, if time is
discrete, this to the engineer represents a time series, such as a speech
signal, or any measurement which depends on time. But we will also allow
functions on lattices such as $\mathbb{Z}^d$.

\quad
         In the case $d = 2$, we may be considering the grayscale numbers
which represent exposure in a digital camera. In this case, the function
(grayscale) is defined on a subset of $\mathbb{Z}^2$, and is then simply a 
matrix.

\quad
         A random walk 
on $\mathbb{Z}^d$ is an assignment of a sequential and random
motion as a function of time. The randomness presupposes assigned
probabilities. But we will use the term ``random walk''
also in connection
with random walks
on combinatorial trees.
\end{quote}

\addtolength{\headsep}{2\baselineskip}
\let\saverunheadstyle\runheadstyle
\def\runheadstyle{\rlap{\raisebox{-2\baselineskip}[0pt][0pt]{\vbox{\begin{minipage}[t]{0.24\textwidth}{\scshape mathematics}\end{minipage}\kern0.01\textwidth\begin{minipage}[t]{0.24\textwidth}{\scshape probability}\end{minipage}\kern0.01\textwidth\begin{minipage}[t]{0.24\textwidth}{\scshape engineering}\end{minipage}\kern0.01\textwidth\begin{minipage}[t]{0.24\textwidth}{\scshape physics}\end{minipage}\smallskip
\hrule}}}\rmfamily\upshape}

\bigskip

\noindent
\begin{minipage}[t]{0.24\textwidth}{\bfseries\raggedright nested\\subspaces}\end{minipage}\kern0.01\textwidth\begin{minipage}[t]{0.24\textwidth}{\bfseries\raggedright refinement}\end{minipage}\kern0.01\textwidth\begin{minipage}[t]{0.24\textwidth}{\bfseries\raggedright multiresolution}%
\index{multiresolution}
\end{minipage}\kern0.01\textwidth\begin{minipage}[t]{0.24\textwidth}{\bfseries\raggedright scales of visual\\resolutions}
\end{minipage}

\begin{quote}
      While finite or infinite families of nested subspaces are ubiquitous
in mathematics, and have been popular in Hilbert space theory for
generations (at least since the 1930s), this idea was revived in
a different guise in 1986 by St\'ephane Mallat, then an engineering graduate
student. In its adaptation to wavelets, 
the idea is now
referred to as the multiresolution method. 

\quad
      What made the idea especially popular in the wavelet 
community was
that it offered a skeleton on which various discrete algorithms in applied
mathematics could be attached and turned into wavelet constructions 
in harmonic analysis.
In fact what we now call multiresolutions\index{multiresolution} have come to signify a
crucial link between the world of discrete wavelet algorithms, which are
popular in computational mathematics and in engineering 
(signal/image processing, data mining,
etc.)\ on the one side, and on the other side continuous wavelet
bases
in function spaces, especially in $L^2(\mathbb{R}^d)$. Further, the
multiresolution idea closely mimics how fractals are analyzed with the use
of finite function systems.

\quad
      But in mathematics, or more precisely in operator theory, 
the
underlying idea dates back to work of John von Neumann, Norbert Wiener, and
Herman Wold, where nested and closed subspaces 
in Hilbert space were used
extensively in an axiomatic approach to stationary processes, especially for
time series. Wold proved that any (stationary) time series can be decomposed
into two different parts: The first (deterministic) part can be exactly
described by a linear combination of its own past, while the second part is
the opposite extreme; it is \emph{unitary}, in the language of von Neumann.

\quad
        von Neumann's version of the same theorem is a pillar in operator
theory. It states that every isometry in a Hilbert space $\mathcal{H}$ is the unique 
sum of a shift isometry and a unitary operator, i.e., the initial Hilbert 
space $\mathcal{H}$ splits canonically as an orthogonal sum of two subspaces $\mathcal{H}_{s}$ 
and $\mathcal{H}_{u}$ in $\mathcal{H}$, one which carries the shift operator, and the other $\mathcal{H}_{u}$ 
the unitary part.  The shift isometry is defined from a nested scale of closed
spaces $V_n$, such that the intersection of these spaces is $\mathcal{H}_{u}$.
Specifically, 
\[
  \cdots \subset V_{-1} \subset V_{0} \subset V_{1} \subset V_{2}
  \subset \cdots  \subset V_{n}  \subset V_{n+1}  \subset \cdots
\]
\[
  \bigwedge_{n}V_{n}=\mathcal{H}_{u}, \text{ and } \bigvee_{n}V_{n}=\mathcal{H}.
\]

\quad
       However, St\'ephane Mallat was motivated instead by the notion of scales
of resolutions in the sense of optics. This in turn is based on a certain 
``artificial-intelligence'' approach to vision and optics, developed earlier by
David Marr at MIT, an approach which imitates the mechanism of vision in the
human eye. 

\quad
       The connection from these developments in the 1980s back
to von Neumann is this: Each of the closed subspaces
$V_n$ corresponds to a
level of resolution
in such a way that a larger subspace represents a finer
resolution.
Resolutions
are relative, not absolute! In this view, the
relative complement of the smaller (or coarser) subspace in larger space
then represents the visual detail which is added in passing from a blurred
image to a finer one, i.e., to a finer visual resolution.

\quad
       This view became an instant hit in the wavelet
\index{wavelet}
community, as it
offered a repository for the fundamental father and the mother functions,
also called the scaling
function $\varphi$, and the wavelet function
$\psi$. Via a
system of translation and scaling operators,
these functions then generate
nested subspaces,
and we recover the scaling
identities which initialize the
appropriate 
\end{quote}

\begin{minipage}[t]{0.24\textwidth}{\scshape mathematics}\end{minipage}\kern0.01\textwidth\begin{minipage}[t]{0.24\textwidth}{\scshape probability}\end{minipage}\kern0.01\textwidth\begin{minipage}[t]{0.24\textwidth}{\scshape engineering}\end{minipage}\kern0.01\textwidth\begin{minipage}[t]{0.24\textwidth}{\scshape physics}\end{minipage}\smallskip

\hrule\medskip

\begin{quote}
algorithms.
What results is now called the family of pyramid
algorithms
in wavelet analysis.
The approach itself is called the
multiresolution approach (MRA) to wavelets.
And in the meantime various
generalizations (GMRAs) have emerged.  
\end{quote}

\begin{quote}
       In all of this, there was a second ``accident'' at play: As it turned
out, pyramid algorithms
in wavelet analysis
now lend 
themselves via
multiresolutions, or nested scales of closed subspaces,
to an analysis based on
frequency bands.
Here we refer 
to bands of frequencies as they have already
been used for a long time in signal processing.

\quad
       One reason for the success in varied disciplines of the same
geometric idea is perhaps that it is closely modeled on how we historically
have represented numbers in the positional number system. 
Analogies to the Euclidean algorithm
seem especially compelling.
\end{quote}

\bigskip

\noindent
\begin{minipage}[t]{0.24\textwidth}{\bfseries\raggedright operator}\end{minipage}\kern0.01\textwidth\begin{minipage}[t]{0.24\textwidth}{\bfseries\raggedright process}\end{minipage}\kern0.01\textwidth\begin{minipage}[t]{0.24\textwidth}{\bfseries\raggedright black box}
\end{minipage}\kern0.01\textwidth\begin{minipage}[t]{0.24\textwidth}{\bfseries\raggedright observable\\(if selfadjoint)}\end{minipage}

\begin{quote} 
          In linear algebra
 students are familiar with the distinctions
between (linear) transformations $T$ (here called ``operators'') and matrices.
For a fixed operator $T \colon V \rightarrow W$, there is a variety of matrices, one for
each choice of basis
 in $V$ and in $W$. In many engineering applications, the
transformations are not restricted to be linear, but instead represent some
experiment (``black box,'' in Norbert Wiener's terminology), one with an input
and an output, usually functions of time. The input could be an external
voltage function, the black box an electric circuit, and the output the
resulting voltage in the circuit. (The output is a solution to a
differential equation.)

\quad
          This context is somewhat different from that of quantum mechanical
(QM) operators $T\colon V \to V$ where $V$ is a Hilbert space. In QM, 
selfadjoint
operators represent observables such as position $Q$ and momentum $P$, or time
and energy.
\end{quote}

\ \bigskip

\noindent
\begin{minipage}[t]{0.24\textwidth}{\bfseries\raggedright Fourier dual\\pair}
\end{minipage}\kern0.01\textwidth\begin{minipage}[t]{0.24\textwidth}{\bfseries\raggedright generating\\function}\end{minipage}\kern0.01\textwidth\begin{minipage}[t]{0.24\textwidth}{\bfseries\raggedright time/frequency}\end{minipage}\kern0.01\textwidth\begin{minipage}[t]{0.24\textwidth}{\bfseries\raggedright $P$/$Q$}\end{minipage}

\begin{quote} 
          The following dual pairs position $Q$/momentum $P$, and time/energy
may be computed with the use of Fourier series or Fourier transforms;
and in
this sense they are examples of Fourier dual
pairs. If for example time is
discrete, then frequency may be represented by numbers in the interval $\left[\,0, 2
\pi\right)$; or in $\left[\,0, 1\right)$ if we enter the number $2 \pi$ into the Fourier exponential.
Functions of the frequency are then periodic, so the two endpoints are
identified. In the case of the interval $\left[\,0, 1\right)$, $0$ on the left is identified
with $1$ on the right. So a low frequency band is an interval centered at $0$,
while a high frequency band is an interval centered at $1/2$.  
\end{quote}

\begin{minipage}[t]{0.24\textwidth}{\scshape mathematics}\end{minipage}\kern0.01\textwidth\begin{minipage}[t]{0.24\textwidth}{\scshape probability}\end{minipage}\kern0.01\textwidth\begin{minipage}[t]{0.24\textwidth}{\scshape engineering}\end{minipage}\kern0.01\textwidth\begin{minipage}[t]{0.24\textwidth}{\scshape physics}\end{minipage}\smallskip

\hrule\medskip

\begin{quote}
\noindent
Let a function $W$
on $\left[\,0, 1\right)$ represent a probability assignment. Such functions $W$
are thought of
as ``filters'' in signal processing.
We say that $W$ is low-pass if it is $1$ at
$0$, or if it is near $1$ for frequencies near $0$. 
Low-pass filters pass signals
with low frequencies, and block the others.

\quad
           If instead some filter $W$ is $1$ at $1/2$, or takes values near $1$
 for
frequencies near $1/2$, then we say that $W$ is high-pass; it passes signals with
high frequency.
\end{quote}

\ \bigskip

\noindent
\begin{minipage}[t]{0.24\textwidth}{\bfseries\raggedright convolution}\end{minipage}\kern0.01\textwidth\begin{minipage}[t]{0.24\textwidth}{\bfseries\raggedright ---}\end{minipage}\kern0.01\textwidth\begin{minipage}[t]{0.24\textwidth}{\bfseries\raggedright filter}\end{minipage}\kern0.01\textwidth\begin{minipage}[t]{0.24\textwidth}{\bfseries\raggedright smearing}\end{minipage}

\begin{quote} 
           Pointwise multiplication of functions of frequencies corresponds
in the Fourier dual
time-domain to the operation of convolution (or of
Cauchy product if the time-scale is discrete.) The process of modifying a
signal with a fixed convolution is called a linear filter in signal
processing.
The corresponding Fourier dual
frequency function is then
referred to as ``frequency response'' or the ``frequency response function.'' 

\quad
           More generally, in the continuous case, since convolution tends
to improve smoothness of functions, physicists call it ``smearing.''
\end{quote} 

\ \bigskip

\noindent
\begin{minipage}[t]{0.24\textwidth}{\bfseries\raggedright decomposition\\(e.g., Fourier\\coefficients
in a\\Fourier \rlap{expansion)}}
\end{minipage}\kern0.01\textwidth\begin{minipage}[t]{0.24\textwidth}{\bfseries\raggedright ---}\end{minipage}\kern0.01\textwidth\begin{minipage}[t]{0.24\textwidth}{\bfseries\raggedright analysis}
\end{minipage}\kern0.01\textwidth\begin{minipage}[t]{0.24\textwidth}{\bfseries\raggedright frequency\\components}\end{minipage}

\begin{quote} 
           Calculating the Fourier coefficients
is ``analysis,''
and adding up
the pure frequencies (i.e., summing the Fourier series)
is called synthesis.
But this view carries over more generally to engineering where there are
more operations involved on the two sides, e.g., breaking up a signal into its
frequency bands, transforming further, and then adding up the ``banded''
functions in the end. If the signal out is the same as the signal in, we say
that the analysis/synthesis
yields perfect reconstruction.
\end{quote}

\bigskip

\noindent
\begin{minipage}[t]{0.24\textwidth}{\bfseries\raggedright integrate\\(e.g., inverse\\Fourier \rlap{transform)}}
\end{minipage}\kern0.01\textwidth\begin{minipage}[t]{0.24\textwidth}{\bfseries\raggedright reconstruct}\end{minipage}\kern0.01\textwidth\begin{minipage}[t]{0.24\textwidth}{\bfseries\raggedright synthesis}\end{minipage}\kern0.01\textwidth\begin{minipage}[t]{0.24\textwidth}{\bfseries\raggedright superposition}\end{minipage}

\begin{quote} 
            Here the terms related to ``synthesis'' refer to the second half
of the kind of signal-processing
design outlined in the previous paragraph.
\end{quote}

\clearpage

\begin{minipage}[t]{0.24\textwidth}{\scshape mathematics}\end{minipage}\kern0.01\textwidth\begin{minipage}[t]{0.24\textwidth}{\scshape probability}\end{minipage}\kern0.01\textwidth\begin{minipage}[t]{0.24\textwidth}{\scshape engineering}\end{minipage}\kern0.01\textwidth\begin{minipage}[t]{0.24\textwidth}{\scshape physics}\end{minipage}\smallskip

\hrule\medskip


\noindent
\begin{minipage}[t]{0.24\textwidth}{\bfseries\raggedright subspace}\end{minipage}\kern0.01\textwidth\begin{minipage}[t]{0.24\textwidth}{\bfseries\raggedright ---}\end{minipage}\kern0.01\textwidth\begin{minipage}[t]{0.24\textwidth}{\bfseries\raggedright resolution}
\end{minipage}\kern0.01\textwidth\begin{minipage}[t]{0.24\textwidth}{\bfseries\raggedright (signals in a)\\frequency band}\end{minipage}

\begin{quote} 
             For a space of functions (signals), the selection of certain
frequencies serves as a way of selecting special signals. When the process
of scaling
is introduced into optics of a digital camera, we note 
that a
nested family of subspaces
corresponds to a grading of visual
resolutions.
\end{quote}

\ \bigskip
\noindent
\begin{minipage}[t]{0.24\textwidth}{\bfseries\raggedright Cuntz relations}
\end{minipage}\kern0.01\textwidth\begin{minipage}[t]{0.24\textwidth}{\bfseries\raggedright ---}\end{minipage}\kern0.01\textwidth\begin{minipage}[t]{0.24\textwidth}{\bfseries\raggedright perfect\\reconstruction\\from subbands}\end{minipage}\kern0.01\textwidth\begin{minipage}[t]{0.24\textwidth}{\bfseries\raggedright subband\\decomposition}\end{minipage}

\begin{quote}
\[
\sum_{i=0}^{N-1} S_{i}S_{i}^{*} = \mathbf{1} \text{, and }
  S_{i}^{*}S_{j} = \delta_{i,j}\mathbf{1}.
\]
\end{quote}

\ \bigskip
\noindent
\begin{minipage}[t]{0.24\textwidth}{\bfseries\raggedright inner product}\end{minipage}\kern0.01\textwidth\begin{minipage}[t]{0.24\textwidth}{\bfseries\raggedright correlation}\end{minipage}\kern0.01\textwidth\begin{minipage}[t]{0.24\textwidth}{\bfseries\raggedright transition\\probability}\end{minipage}\kern0.01\textwidth\begin{minipage}[t]{0.24\textwidth}{\bfseries\raggedright probability\\of transition\\from one state\\to another}\end{minipage}

\begin{quote} 
             In many applications, a vector space with inner product
captures perfectly the geometric and probabilistic features of the
situation. This can be axiomatized in the language of Hilbert space;
and the
inner product is the most crucial ingredient in the familiar axiom system
for Hilbert
space.
\end{quote}

\ \bigskip

\noindent
\begin{minipage}[t]{0.24\textwidth}{\bfseries\raggedright $f_{\operatorname{out}} = T f_{\operatorname{in}}$}\end{minipage}\kern0.01\textwidth\begin{minipage}[t]{0.24\textwidth}{\bfseries\raggedright ---}\end{minipage}\kern0.01\textwidth\begin{minipage}[t]{0.24\textwidth}{\bfseries\raggedright input/output}\end{minipage}\kern0.01\textwidth\begin{minipage}[t]{0.24\textwidth}{\bfseries\raggedright transformation\\of states}
\end{minipage}

\begin{quote} 
             Systems theory language for operators $T \colon V \rightarrow W$. Then vectors
in $V$ are input, and in the range of $T$ output.
\end{quote}

\ \bigskip

\noindent
\begin{minipage}[t]{0.24\textwidth}{\bfseries\raggedright fractal}\end{minipage}\kern0.01\textwidth\begin{minipage}[t]{0.24\textwidth}{\bfseries\raggedright ---}\end{minipage}\kern0.01\textwidth\begin{minipage}[t]{0.24\textwidth}{\bfseries\raggedright ---}\end{minipage}\kern0.01\textwidth\begin{minipage}[t]{0.24\textwidth}{\bfseries\raggedright ---}\end{minipage}

\begin{quote} 
       Intuitively, think of a fractal as reflecting similarity of scales
such as is seen in fern-like images that look ``roughly'' the same at small
and at large scales. Fractals are produced from an infinite iteration of a
finite set of maps, and this algorithm is perfectly suited to the kind of
subdivision which is a cornerstone of the discrete wavelet algorithm.  
Self-similarity could refer
alternately to space, and to time. And further versatility is added, in
that flexibility is allowed into the definition of ``similar.''
\end{quote}

\noindent
\begin{minipage}[t]{0.24\textwidth}{\bfseries\raggedright ---}\end{minipage}\kern0.01\textwidth\begin{minipage}[t]{0.24\textwidth}{\bfseries\raggedright ---}\end{minipage}\kern0.01\textwidth\begin{minipage}[t]{0.24\textwidth}{\bfseries\raggedright data mining}%
\end{minipage}\kern0.01\textwidth\begin{minipage}[t]{0.24\textwidth}{\bfseries\raggedright ---}\end{minipage}

\begin{quote}
The problem of 
how to handle and make use of
 large volumes of data is a corollary of
the digital revolution. As a result, the subject of data mining
itself
changes rapidly. Digitized information (data) is now easy to capture automatically and
to store electronically.
  In science, in commerce, and in industry, data
represents collected observations and information:  In business, there is
data on markets, competitors, and customers.
In manufacturing, there is data
for optimizing production opportunities, and for improving processes.
A tremendous potential for data mining
exists in medicine, genetics, and energy.
But
raw data is not always directly usable, as is evident by inspection.  A key to
advances is our ability to \emph{extract information and knowledge} from the data 
(hence ``data mining''),
and to understand the phenomena governing data sources. Data
mining
is now taught in a variety of forms in engineering departments, as
well as in statistics and computer science departments.

\quad
One of the structures often hidden in data sets is some degree of \emph{scale}.
The goal is to detect and identify one or more natural global and local
scales in the data. Once this is done, it is often possible to
detect associated similarities of scale, much like the familiar
scale-similarity from multidimensional wavelets,
and from fractals. Indeed,
various adaptations of wavelet-like algorithms
 have been shown to be useful.
These algorithms
 themselves are useful in \emph{detecting}
scale-similarities, and are applicable to other types of pattern recognition. Hence,
in this context, generalized multiresolutions
offer another tool for
discovering structures in large data sets, such as those stored in the resources of
the Internet. Because of the sheer volume of data involved, a strictly 
manual analysis is out of the question. Instead, sophisticated query
processors based on statistical and mathematical techniques are used in
generating insights and extracting conclusions from data sets. 
\end{quote}

\subsubsection*{Multiresolutions}
    Haar's
work in 1909--1910 had
    implicitly the key idea which got wavelet
mathematics started on a
    roll 75 years later with Yves Meyer, Ingrid Daubechies, St\'ephane Mallat,
    and others---namely the idea of a multiresolution.\index{multiresolution}
In that respect Haar
was ahead of his time.  
See Figures 1 and 2 for details.



The word ``multiresolution''
suggests a
connection to optics from physics. So that should have been a hint to
mathematicians to take a closer look at trends in signal and image
processing!
Moreover, even staying within mathematics, it turns out that as a
general notion this same idea of a ``multiresolution''
has long roots in
mathematics, even in such modern and pure areas as operator theory
and Hilbert-space
geometry. 
Looking even closer at these interconnections, we can
now recognize scales of subspaces (so-called multiresolutions)
in classical
algorithmic
 construction of orthogonal bases
 in inner-product spaces, now
taught in lots of mathematics courses under the name of the Gram--Schmidt algorithm.
Indeed, a closer look at good old Gram--Schmidt
reveals that it is a matrix
algorithm,
Hence new mathematical tools involving non-com\-mu\-ta\-tiv\-ity!

       If the signal to be analyzed
 is an image, then why not select a fixed
but suitable \emph{resolution}
(or a subspace of signals corresponding to a
selected resolution), 
and then do the computations there? The
selection of a fixed ``resolution''
is dictated by practical concerns. That
idea was key in turning computation of wavelet coefficients
into iterated
matrix algorithms.
As the matrix operations get large, the computation is
carried out in a variety of paths arising from big matrix products.
The dichotomy, continuous vs.\ discrete, is
quite familiar to engineers. The industrial engineers typically work with
huge volumes of numbers.

        Numbers! --- So why wavelets?
Well, what matters to the industrial
engineer is not really the wavelets,
but the fact that special wavelet
functions serve as an efficient way to encode large data sets---I mean encode
for computations. And the wavelet algorithms
 are computational. They work on
numbers. Encoding numbers into pictures, images, or graphs of functions
comes later, perhaps at the very end of the computation. But without the
graphics, I doubt that we would understand any of this half as well as we do
now. The same can be said for the many issues that relate to the crucial
mathematical concept of self-similarity, as we know it from fractals, and
more generally from recursive algorithms.

\section{Definition}
\label{sec:Over}
In this paper we outline several points of view on the interplay between discrete and continuous wavelet transforms; stressing both pure and applied aspects of both. We outline some new links between the two transform technologies based on the theory of representations of generators and relations. By this we mean a finite system of generators which are represented by operators in Hilbert space. We further outline how these representations yield sub-band filter banks for signal and image processing algorithms.

The word ``wavelet transform" (WT) means different things to different people: 
Pure and applied mathematicians typically give different answers the 
questions ``What is the WT?"  And engineers in turn have their own preferred 
quite different approach to WTs.  Still there are two main trends in how WTs
are used, the \emph{continuous} WT on one side, and the \emph{discrete} WT on
the other. Here we offer a userfriendly outline of both, but with a slant
toward geometric methods from the theory of operators in Hilbert space.

Our paper is organized as follows: For the benefit of diverse reader groups, we begin
with Glossary (section \ref{sec2:Glo}). This is a substantial part of our 
account, and it reflects the multiplicity of how the subject is used.  

The concept of multiresolutions or multiresolution analysis (MRA) serves as a
link between the discrete and continuous theory.  

In section \ref{sec3:List}, we summarize how different mathematicians and 
scientists have contributed to and shaped the subject over the years. 

The next two sections then offer a technical overview of both discrete and the 
continuous WTs.  This includes basic tools from Fourier analysis and from 
operators in Hilbert space.  In sections \ref{sec6:Tools} and \ref{sec7:TO} we 
outline the connections between the separate parts of mathematics and their
applications to WTs.    

\section{Introduction}
\label{sec1:Intro}
While applied problems such as time series, signals and processing of digital 
images come from engineering and from the sciences, they have in the past two 
decades taken a life of their own as an exciting new area of applied 
mathematics. While searches in Google on these keywords typically yield sites 
numbered in the millions, the diversity of applications is wide, and it seems 
reasonable here to narrow our focus to some of the approaches that are both 
more mathematical and more recent. For references, see for example 
\cite{AuKo06, BrMa06, Liu06, StNg96}.  In addition, our own interests (e.g., 
\cite{Jor03, Jor06a, Son05,Son06}) have colored the presentation below. Each of the 
two areas, the discrete side, and the continuous theory is huge as measured by 
recent journal publications. A leading theme in our article is the independent 
interest in a multitude of interconnections between the discrete algorithm and 
their uses in the more mathematical analysis of function spaces (continuous 
wavelet transforms). The mathematics involved in the study and the applications 
of this interaction we feel is of benefit to both mathematicians and to 
engineers. See also \cite{Jor03}. An early paper \cite{DaLa92} by Daubechies 
and Lagarias was especially influential in connecting the two worlds, discrete 
and continuous.

\section{The discrete vs continuous wavelet Algorithms}
\label{sec5:Cont}

\subsection{The Discrete Wavelet Transform}
If one stays with function spaces, it is then popular to pick the 
$d$-dimensional Lebesgue measure on $\br^{d}$, $d = 1, 2, ¡¦$, and pass to the 
Hilbert space $L^{2}(\br^{d})$ of all square integrable functions on 
$\br^{d}$, referring to d-dimensional Lebesgue measure. A wavelet basis refers 
to a family of basis functions for $L^{2}(\br^{d})$ generated from a finite 
set of normalized functions $\psi_{i}$ , the index $i$  chosen from a fixed 
and finite index set $I$, and from two operations, one called scaling, and the 
other translation. The scaling is typically specified by a $d$ by $d$ matrix 
over the integers $\bz$ such that all the eigenvalues in modulus are bigger 
than one, lie outside the closed unit disk in the complex plane. The 
$d$-lattice is denoted $\mathbb{Z}^{d}$ , and the translations will be by vectors 
selected from $\mathbb{Z}^{d}$. We say that we have a wavelet basis if the 
triple indexed family
$\psi_{i,j,k}(x) := | det A |^{j/2} \psi( A^{j} x + k)$ forms an orthonormal 
basis (ONB) for $L^{2}(\br^{d})$ as $i$ varies in $I$, $j \in \bz$, and 
$k \in \br^{d}$. The word ``orthonormal" for a family $F$ of vectors in a 
Hilbert space $\mathcal{H}$ refers to the norm and the inner product in 
$\mathcal{H}$:  The vectors in an orthonormal family F are assumed to have 
norm one, and to be mutually orthogonal. If the family is also total (i.e., 
the vectors in $F$ span a subspace which is dense in $\mathcal{H}$), we say 
that $F$ is an orthonormal basis (ONB.) 

While there are other popular wavelet bases, for example frame bases, and dual 
bases (see e.g., \cite{BJMP05, DuRo07a} and the papers cited there), the ONBs 
are the most agreeable at least from the mathematical point of view. 

That there are bases of this kind is not at all clear, and the subject of 
wavelets in this continuous context has gained much from its connections to 
the discrete world of signal- and image processing.

Here we shall outline some of these connections with an emphasis on the 
mathematical context. So we will be stressing the theory of Hilbert space, and 
bounded linear operators acting in Hilbert space $\mathcal{H}$, both 
individual operators, and families of operators which form algebras. 

As was noticed recently the operators which specify particular subband 
algorithms from the discrete world of signal- processing turn out to satisfy 
relations that were found (or rediscovered independently) in the theory of 
operator algebras, and which go under the name of Cuntz algebras, denoted 
$\mathcal{O}_{N}$ if $n$ is the number of bands. For additional details, see 
e.g., \cite{Jor06a}.

In symbols the $C^{*}-$algebra has generators $(S_{i})_{i=0}^{N-1}$, and the
relations are
\begin{equation}
\label{E:Si}
\sum_{i=0}^{N-1} S_{i}S_{i}^{*} = \mathbf{1}
\end{equation}
(where $\mathbf{1}$ is the identity element in $\mathcal{O}_{N}$) and
\begin{equation}
\label{E:Sdelta}
\sum_{i=0}^{N-1} S_{i}S_{i}^{*} = \mathbf{1} \text{, and }
  S_{i}^{*}S_{j} = \delta_{i,j}\mathbf{1}.
\end{equation}

In a representation on a Hilbert space, say $\mathcal{H}$, the symbols $S_{i}$ 
turn into bounded operators, also denoted $S_{i}$, and the identity element 
$\mathbf{1}$ turns into the identity operator $I$ in $\mathcal{H}$, i.e., the 
operator $I:h \to h$, for $h \in \mathcal{H}$.  In operator language, the two 
formulas \ref{E:Si} and \ref{E:Sdelta} state that each $S_{i}$ is an isometry 
in $\mathcal{H}$, and that te respective ranges $S_{i}\mathcal{H}$ are 
mutually orthogonal, i.e., $S_{i}\mathcal{H} \perp S_{j}\mathcal{H}$ for 
$i \ne j$.  Introducing the projections $P_{i} = S_{i}S_{i}^{*}$, we get 
$P_{i}P_{j} = \delta_{i,j}P_{i}$, and 
\[  
  \sum_{i=0}^{N-1} P_{i} = I
\]
In the engineering literature this takes the form of programming diagrams:


If the process of Figure 3 
is repeated, we arrive at the discrete 
wavelet transform


or stated in the form of images ($n=5$)



Selecting a resolution subspace 
$V_{0} = closure$ $span\{\varphi(\cdot -k)|k \in \mathbb{Z}\}$, we arrive at a 
wavelet subdivision $\{\psi_{j,k}|j \geq 0, k \in \mathbb{Z}\}$, where 
$\psi_{j,k}(x) = 2^{j/2} \psi(2^{j}x-k)$, and the continuous expansion 
$f = \sum_{j,k} \langle \psi_{j,k}|f \rangle \psi_{j,k}$ or the discrete 
analogue derived from the isometries, $i=1,2, \cdots, N-1$, $S_{0}^{k}S_{i}$ 
for $k=0,1,2, \cdots$; called the discrete wavelet transform.

\subsubsection*{Notational convention.} 
In algorithms, the letter $N$ is 
popular, and often used for counting more than one thing.

In the present contest of the Discete Wavelet Algorithm (DWA) or DWT,  we 
count two things, ``the number of times a picture is decomposed via 
subdivision". We have used $n$ for this. The other related but different 
number $N$ is the number of subbands, $N = 2$ for the dyadic DWT, and $N = 4$ 
for the image DWT. The image-processing WT in our present context is the 
tensor product of the 1-D dyadic WT, so $2 \times 2 = 4$. Caution: Not all 
DWAs arise as tensor products of $N = 2$ models. The wavelets coming from 
tensor products are called separable. When a particular image-processing 
scheme is used for generating continuous wavelets it is not transparent if we 
are looking at a separable or inseparable wavelet!

To clarify the distinction, it is helpful to look at the representations of 
the Cuntz relations by operators in Hilbert space. We are dealing with 
representations of the two distinct algebras $\mathcal{O}_{2}$, and 
$\mathcal{O}_{4}$; two frequency subbands vs 4 subbands. Note that the Cuntz 
$\mathcal{O}_{2}$, and $\mathcal{O}_{4}$ are given axiomatic, or purely 
symbolically. It is only when subband filters are chosen that we get 
representations. This also means that the choice of $N$ is made initially; and 
the same $N$ is used in different runs of the programs. In contrast, the 
number of times a picture is decomposed varies from one experiment to the next!

\textbf{Summary:} $N = 2$ for the dyadic DWT: The operators in the 
representation are $S_{0}$ , $S_{1}$. One average operator, and one detail 
operator. The detail operator $S_{1}$ ``counts"  local detail variations.

Image-processing. Then $N = 4$ is fixed as we run different images in the 
DWT: The operators are now: $S_{0}$ , $S_{H}$, $S_{V}$, $S_{D}$. One average 
operator, and three detail operator for local detail variations in the three 
directions in the plane.

\subsection{The Continuous Wavelet Transform}
Consider functions $f$ on the real line $\mathbb{R}$.  We select the Hilbert
space of functions to be $L^{2}(\mathbb{R})$ To start a continuous WT, we
must select a function $\psi \in L^{2}(\mathbb{R})$ and $r, s \in \mathbb{R}$
such that the following family of functions 
\[
  \psi_{r,s}(x) = r^{-1/2}\psi(\frac{x-s}{r})
\]
creates an over-complete basis for $L^{2}(\mathbb{R})$.  An over-complete
family of vectors in a Hilbert space is often called a coherent decomposition.
This terminology comes from quantum optics.  What is needed for a continuous 
WT in the simplest case is the following representation valid for all 
$f \in L^{2}(\mathbb{R})$:
\[
  f(x)=C_{\psi}^{-1}\int \int_{\mathbb{R}^{2}} \langle \psi_{r,s}|f \rangle \psi_{r,s}(x) \frac{drds}{r^{2}}
\]
where
$C_{\psi} := \int_{\mathbb{R}}|\hat{\psi}(\omega)|^{2}\frac{d\omega}{\omega}$
and where $\langle \psi_{r,s}|f \rangle=\int_{\mathbb{R}}\overline{\psi_{r,s}(y)}f(y)dy$.
The refinements and implications of this are spelled out in tables in section 
\ref{sec:Conn}

\subsection{Some background on Hilbert space}

Wavelet theory is the art of finding a special kind of basis in Hilbert space.
Let $\mathcal{H}$ be a Hilbert space over $\mathbb{C}$ and denote the inner
product $\left\langle \,\cdot\mid\cdot\,\right\rangle $. For us, it
is assumed linear in the second variable. If $\mathcal{H}=L^{2}\left(
\mathbb{R}\right)  $, then
\[
\left\langle \,f\mid g\,\right\rangle :=\int_{\mathbb{R}}\overline{f\left(
x\right)  }\,g\left(  x\right)  \,dx.
\]
If $\mathcal{H}=\ell^{2}\left(  \mathbb{Z}\right)  $, then
\[
\left\langle \,\xi\mid\eta\,\right\rangle :=\sum_{n\in\mathbb{Z}}\bar{\xi}_{n}
\eta_{n}.
\]
Let $\mathbb{T}=\mathbb{R}/2\pi\mathbb{Z}$. If $\mathcal{H}=L^{2}\left(
\mathbb{T}\right)  $, then
\[
\left\langle \,f\mid g\,\right\rangle :=\frac{1}{2\pi}\int_{-\pi}^{\pi
}\overline{f\left(  \theta\right)  }\,g\left(  \theta\right)  \,d\theta
.
\]
Functions $f\in L^{2}\left(  \mathbb{T}\right)  $ have Fourier series: Setting
$e_{n}\left(  \theta\right)  =e^{in\theta}$,
\[
\hat{f}\left(  n\right)  :=\left\langle \,e_{n}\mid f\,\right\rangle =\frac
{1}{2\pi}\int_{-\pi}^{\pi}e^{-in\theta}f\left(  \theta\right)  \,d\theta
,
\]
and
\[
\left\Vert f\right\Vert _{L^{2}\left(  \mathbb{T}\right)  }^{2}=\sum
_{n\in\mathbb{Z}}\left\vert \hat{f}\left(  n\right)  \right\vert
^{2}.
\]
Similarly if $f\in L^{2}\left(  \mathbb{R}\right)  $, then
\[
\hat{f}\left(  t\right)  :=\int_{\mathbb{R}}e^{-ixt}f\left(  x\right)
\,dx,
\]
and
\[
\left\Vert f\right\Vert _{L^{2}\left(  \mathbb{R}\right)  }^{2}=\frac{1}{2\pi
}\int_{\mathbb{R}}\left\vert \hat{f}\left(  t\right)  \right\vert
^{2}\,dt.
\]

Let $J$ be an index set. We shall only need to consider the case when $J$ is
countable. Let $\left\{  \psi_{\alpha}\right\}  _{\alpha\in J}$ be a family of
nonzero vectors in a Hilbert space $\mathcal{H}$. We say it is an
\emph{orthonormal basis} (ONB) if
\begin{equation}
\left\langle \,\psi_{\alpha}\mid\psi_{\beta}\,\right\rangle =\delta
_{\alpha,\beta}\text{\qquad(Kronecker delta)}\label{eq1.2.8}
\end{equation}
and if
\begin{equation}
\sum_{\alpha\in J}\left\vert \left\langle \,\psi_{\alpha}\mid f\,\right\rangle
\right\vert ^{2}=\left\Vert f\right\Vert ^{2}\text{\qquad holds for all }
f\in\mathcal{H}.\label{eq1.2.9}
\end{equation}
If only (\ref{eq1.2.9}) is assumed, but not (\ref{eq1.2.8}), we say that
$\left\{  \psi_{\alpha}\right\}  _{\alpha\in J}$ is a (normalized) \emph{tight
frame}. We say that it is a \emph{frame} with \emph{frame constants} $0<A\leq
B<\infty$ if
\[
A\left\Vert f\right\Vert ^{2}\leq\sum_{\alpha\in J}\left\vert \left\langle
\,\psi_{\alpha}\mid f\,\right\rangle \right\vert ^{2}\leq B\left\Vert
f\right\Vert ^{2}\text{\qquad holds for all }f\in\mathcal{H}.
\]
Introducing the rank-one operators $Q_{\alpha}:=\left\vert \psi_{\alpha
}\right\rangle \left\langle \psi_{\alpha}\right\vert $ of Dirac's terminology,
see \cite{BrJo02}, we see that $\left\{  \psi_{\alpha}\right\}  _{\alpha\in
J}$ is an ONB if and only if the $Q_{\alpha}$'s are projections, and
\begin{equation}
\sum_{\alpha\in J}Q_{\alpha}=I\qquad(=\text{the identity operator in
}\mathcal{H}).\label{eq1.2.10}
\end{equation}
It is a (normalized) tight frame if and only if (\ref{eq1.2.10}) holds but
with no further restriction on the rank-one operators $Q_{\alpha}$. It is a
frame with frame constants $A$ and $B$ if the operator
\[
S:=\sum_{\alpha\in J}Q_{\alpha}
\]
satisfies
\[
AI\leq S\leq BI
\]
in the order of hermitian operators. (We say that operators $H_{i}=H_{i}
^{\ast}$, $i=1,2$, satisfy $H_{1}\leq H_{2}$ if $\left\langle
\,f\mid H_{1}f\,\right\rangle \leq\left\langle \,f\mid H_{2}f\,\right\rangle $
holds for all $f\in\mathcal{H}$). If $h,k$ are vectors in a Hilbert
space $\mathcal{H}$, then the operator
$A=\left| h\right\rangle \left\langle k\right| $ is defined by the identity
$\left\langle \,u\mid Av\,\right\rangle =\left\langle \,u\mid h\,\right\rangle \left\langle \,k\mid v\,\right\rangle $
for all $u,v\in\mathcal{H}$.

Wavelets in $L^{2}\left(  \mathbb{R}\right)  $ are generated by simple
operations on one or more functions $\psi$ in $L^{2}\left(  \mathbb{R}\right)
$, the operations come in pairs, say scaling and translation, or
phase-modulation and translations. If $N\in\left\{  2,3,\dots\right\}  $ we
set
\[
\psi_{j,k}\left(  x\right)  :=N^{j/2}\psi\left(  N^{j}x-k\right)  \text{\qquad
for }j,k\in\mathbb{Z}.
\]

\subsubsection{Increasing the dimension}
In wavelet theory, \cite{Dau92} there is a tradition for reserving
$\varphi$ for the father function and $\psi$ for the mother function.
A 1-level wavelet transform of an $N \times M$ image can be represented as
\begin{equation}
\label{E:ahvd1}
  \mathbf{f} \mapsto
  \begin{pmatrix}
    \mathbf{a}^{1} & | & \mathbf{h}^{1} \\
    -- & & -- \\
    \mathbf{v}^{1} & | & \mathbf{d}^{1}
  \end{pmatrix} \\
\end{equation}
where the subimages $\mathbf{h}^{1}, \mathbf{d}^{1}, \mathbf{a}^{1}$ and
$\mathbf{v}^{1}$ each have the dimension of $N/2$ by $M/2$.

\begin{equation}
\label{E:ahvdrel}
  \begin{array}{l}
  \mathbf{a}^{1} = V_{m}^{1} \otimes V_{n}^{1} : \varphi^{A}(x,y) = \varphi(x)\varphi(y) 
  = \sum_{i}\sum_{j}h_{i}h_{j}\varphi(2x-i)\varphi(2y-j) \\
  \mathbf{h}^{1} = V_{m}^{1} \otimes W_{n}^{1} : \psi^{H}(x,y) = \psi(x)\varphi(y)
  = \sum_{i}\sum_{j}g_{i}h_{j}\varphi(2x-i)\varphi(2y-j) \\
  \mathbf{v}^{1} = W_{m}^{1} \otimes V_{n}^{1} : \psi^{V}(x,y) = \varphi(x)\psi(y)
  = \sum_{i}\sum_{j}h_{i}g_{j}\varphi(2x-i)\varphi(2y-j) \\
  \mathbf{d}^{1} = W_{m}^{1} \otimes W_{n}^{1} : \psi^{D}(x,y) = \psi(x)\psi(y)
  = \sum_{i}\sum_{j}g_{i}g_{j}\varphi(2x-i)\varphi(2y-j)
  \end{array}
\end{equation}
where $\varphi$ is the father function and $\psi$ is the mother function in 
sense of wavelet, $V$ space denotes the average space and the $W$ spaces are 
the difference space from multiresolution analysis (MRA) \cite{Dau92}.  

In the formulas, we have the following two indexed number systems 
$\mathbf{a}:=(h_{i})$ and $\mathbf{d}:=(g_{i})$, $\mathbf{a}$ is for averages, 
and $\mathbf{d}$ is for local differences.  They are really the input for the
DWT. But they also are the key link between the two transforms, the discrete
and continuous. The link is made up of the following scaling identities:
\[
  \varphi(x)=2\sum_{i \in \mathbb{Z}} h_{i}\varphi(2x-i);
\]
\[
  \psi(x)=2\sum_{i \in \mathbb{Z}} g_{i}\varphi(2x-i);
\]
and (low-pass normalization) $\sum_{i \in \mathbb{Z}} h_{i} = 1$. 
The scalars $(h_{i})$ may be real or complex; they may be finite or infinite 
in number.  If there are four of them, it is called the ``four tap", etc.  The
finite case is best for computations since it corresponds to compactly 
supported functions.  This means that the two functions $\varphi$ and $\psi$ 
will vanish outside some finite interval on a real line.  

The two number systems are further subjected to orthgonality relations, of
which 
\begin{equation}
\label{E:hsum}
  \sum_{i \in \mathbb{Z}} \bar{h}_{i}h_{i+2k} = \frac{1}{2}\delta_{0,k} 
\end{equation}
is the best known.

The systems $h$ and $g$ are both low-pass and high-pass filter coefficients.
In \ref{E:ahvdrel}, $\mathbf{a}^{1}$ denotes the first averaged image, which consists of 
average intensity values of the original image.  Note that only $\varphi$ 
function, $V$ space and $h$ coefficients are used here. Similarly,
$\mathbf{h}^{1}$ denotes the first detail image of horizontal components, which consists of intensity difference along the vertical axis of the original 
image. Note that $\varphi$ function is used on $y$ and $\psi$ function on 
$x$, $W$ space for $x$ values and $V$ space for $y$ values; and both $h$ and 
$g$ coefficients are used accordingly.  The data 
$\mathbf{v}^{1}$ denotes the first detail image of vertical components, which 
consists of intensity difference along the horizontal axis of the original 
image. Note that $\varphi$ function is used on $x$ and $\psi$ function on 
$y$, $W$ space for $y$ values and $V$ space for $x$ values; and both $h$ and 
$g$ coefficients are used accordingly. Finally, 
$\mathbf{d}^{1}$ denotes the first detail image of diagonal components, which consists of intensity difference along the diagonal axis of the original image. The original image is reconstructed from the decomposed image by taking the sum 
of the averaged image and the detail images and scaling by a scaling factor. It
could be noted that only $\psi$ function, $W$ space and $g$ coefficients are 
used here. 
See \cite{Wal99, Son06}.

This decomposition not only limits to one step but it can be done again and 
again on the averaged detail depending on the size of the image.  Once it 
stops at certain level, quantization (see \cite{Sko01, Use01}) is done on the 
image.  This quantization step may be lossy or 
lossless.  Then the lossless entropy encoding is done on 
the decomposed and quantized image.  

The relevance of the system of identities (\ref{E:hsum}) may be summarized as
follows.  Set 
\[
  m_{0}(z):= \frac{1}{2} \sum_{k \in \mathbb{Z}}h_{k}z^{k} \text{ for all } 
  z \in \mathbb{T}; 
\]
\[
  g_{k}:= (-1)^{k}\bar{h}_{1-k} \text{ for all } k \in \mathbb{Z}; 
\]
\[
  m_{1}(z):= \frac{1}{2} \sum_{k \in \mathbb{Z}}g_{k}z^{k}; \text{ and }
\] 
\[ 
  (S_{j}f)(z)=\sqrt{2}m_{j}(z)f(z^{2}), \text{ for } j=0,1,  \text{ } 
  f \in L^{2}(\mathbb{T}), \text{ } z \in \mathbb{T}.
\]
Then the following conditions are equivalent:
\begin{itemize}
\item[(a)] The system of equations (\ref{E:hsum}) is satisfied.
\item[(b)] The operators $S_{0}$ and $S_{1}$ satisfy the Cuntz relations.
\item[(c)] We have perfect reconstruction in the subband system of Figure 3.
\end{itemize}

Note that the two operators $S_{0}$ and $S_{1}$ have equivalent matrix 
representations.  Recall that by Parseval's formula we have 
$L^{2}(\mathbb{T}) \simeq l^{2}(\mathbb{Z})$.  So representing $S_{0}$ instead 
as an $\infty \times \infty$ matrix acting on column vectors 
$x =(x_{j})_{j \in \mathbb{Z}}$ we get 
\[
  (S_{0}x)_{i} = \sqrt{2} \sum_{j\in \mathbb{Z}} h_{i-2j} x_{j}
\]
and for the adjoint operator $F_{0} := S_{0}^{*}$, we get the matrix 
representation
\[
  (F_{0}x)_{i}=\frac{1}{\sqrt{2}}\sum_{j \in \mathbb{Z}} \bar{h}_{j-2i} x_{j}
\]
with the overbar signifying complex conjugation.  This is computational 
significance to the two matrix representations, both the matrix for $S_{0}$,
and for $F_{0} := S_{0}^{*}$, is slanted.  However, the slanting of one is the
mirror-image of the other, i.e., 


\subsubsection*{Significance of slanting}
The slanted matrix representations refers to the corresponding operators in 
$L^{2}$. In general operators in Hilbert function spaces have many matrix 
representations, one for each orthonormal basis (ONB), but here we are 
concerned with the ONB consisting of the Fourier frequencies $z^{j}$, 
$j \in \mathbb{Z}$. So in our matrix representations for the $S$ operators and their 
adjoints we will be acting on column vectors, each infinite column 
representing a vector in the sequence space $l^{2}$. A vector in $l^{2}$ is 
said to be of finite size if it has only a finite set of non-zero entries.

    It is the matrix $F_{0}$ that is effective for iterated matrix 
computation. Reason: When a column vector $x$ of a fixed size, say 2 s is 
multiplied, or acted on by $F_{0}$,  the result is a vector $y$ of half the 
size, i.e., of size $s$. So $y = F_{0} x$. If we use $F_{0}$ and $F_{1}$ 
together on $x$, then we get two vectors, each of size $s$, the other one 
$z = F_{1} x$, and we can form the combined column vector of $y$ and $z$; 
stacking $y$ on top of $z$. In our application, $y$ represents averages, while 
$z$ represents local differences: Hence the wavelet algorithm.

\[
  \begin{bmatrix}
    \vdots \\ y_{-1} \\ y_{0} \\ y_{1} \\ \vdots \\ -- \\ \vdots \\ 
    z_{-1} \\ z_{0} \\ z_{1} \\ \vdots
  \end{bmatrix}
  =
  \begin{bmatrix}
    F_{0} \\ -- \\ F_{1}
  \end{bmatrix}
  \begin{bmatrix}
    \vdots \\ x_{-2} \\ x_{-1} \\ x_{0} \\ x_{1} \\ x_{2} \\ \vdots
  \end{bmatrix}
\]

\[
  y=F_{0}x
\]

\[
  z=F_{1}x
\]

\subsection{Connections to group theory}
\label{sec:Conn}

The first line in the two tables
below
is the continuous wavelet transform. 
It comes from what in physics is called 
\emph{coherent vector
decompositions}.
Both transforms applies to vectors in Hilbert space $\mathcal{H}$, and 
$\mathcal{H}$ may vary from case to case.  Common to all transforms is vector 
input and output.  If the input agrees with output we say that the combined 
process yields the identity operator image.  
$\mathbf{1}: \mathcal{H} \to \mathcal{H}$ or written 
$\mathbf{1}_{\mathcal{H}}$.  So for example if $(S_{i})_{i=0}^{N-1}$ is a 
finite operator system, and input/output operator example may take the form 
\[  
\sum_{i=0}^{N-1}S_{i}S_{i}^{*}=\mathbf{1}_{\mathcal{H}}.
\]

Summary of and variations on the resolution
of the identity operator $\mathbf{1}$ in $L^{2}$ or in $\ell^{2}$, for 
$\psi$ and
$\tilde{\psi}$ where 
$\psi_{r,s}\left( x\right) 
=r^{-\frac{1}{2}}\psi\left( \frac{x-s}{r}\right) $,

\[
C_{\psi}=\int_{\mathbb{R}}\frac{d\omega}{\left| \omega\right| }
\left| \smash{\hat{\psi}}\left( \omega\right) \right| ^{2}<\infty,
\]
similarly for $\tilde{\psi}$ and 
$C_{\psi,\tilde{\psi}}=\int_{\mathbb{R}}\frac{d\omega}{\left| \omega\right| }
\overline{\hat{\psi}\left( \omega\right) }\,
\Hat{\Tilde{\psi}}\left( \omega\right) $:

\addvspace{\bigskipamount}
\noindent
$\renewcommand{\arraystretch}{1.25}
\begin{tabular}
[c]
{p{0.15\textwidth}
||p{0.375\textwidth}
|p{0.375\textwidth}}
\multicolumn{3}{c}{\parbox{0.9\textwidth}
{\rule[-12pt]{0pt}{12pt}}}\\
$N=2$ & Overcomplete Basis & Dual Bases \\\hline\hline
continu\-ous resolution
& $\displaystyle C_{\psi}^{-1}\!\!\iint\limits_{\mathbb{R}^{2}}^{{}}
\frac{dr\,ds}{r^{2}}
\left| \psi_{r,s}\right\rangle \!\left\langle \psi_{r,s}\right| 
$\par
\hfill
$\displaystyle {}
=\mathbf{1}\vphantom{\iint}$ 
& $\displaystyle 
C_{\!\psi,\tilde{\psi}}^{-1}\!\iint\limits_{\mathbb{R}^{2}}^{{}}
\frac{dr\,ds}{r^{2}}
\left| \psi_{r,s}\right\rangle 
\!\left\langle \smash{\tilde{\psi}_{r,s}}\right| \vphantom{\tilde{\psi}_{r,s}}
$\par
\hfill
$\displaystyle {}
=\mathbf{1}\vphantom{\iint}$ \\\hline
discrete resolution
& {\raggedright $\displaystyle 
\sum_{\vphantom{k}j\in\mathbb{Z}}\sum_{\vphantom{j}k\in\mathbb{Z}}
\left| \psi_{j,k}\right\rangle \left\langle \psi_{j,k}\right| 
=\mathbf{1}$\,, \\
$\psi_{j,k}$ corresponding to \\
$r=2^{-j}$, 
$s=k2^{-j}$}
& $\displaystyle 
\sum_{\vphantom{k}j\in\mathbb{Z}}\sum_{\vphantom{j}k\in\mathbb{Z}}
\left| \psi_{j,k}\right\rangle 
\left\langle \smash{\tilde{\psi}_{j,k}}\right| \vphantom{\tilde{\psi}_{j,k}}
=\mathbf{1}$ 
\\\hline\hline
$N\geq 2$ 
& Isometries in $\ell^{2}$ 
& Dual Operator System in $\ell^{2}$ \\\hline\hline
sequence spaces 
& {\raggedright $\displaystyle \sum_{i=0_{\mathstrut}}^{N-1^{\mathstrut}}
S_{i}S_{i}^{\ast}=\mathbf{1}$\,, \\
where
$\displaystyle S_{0},\dots ,S_{N-1}$ \\
are adjoints to the \\
quadrature mirror filter \\
operators $F_{i}$, i.e., 
$\displaystyle S_{i}=F_{i}^{\ast}$} 
& {\raggedright $\displaystyle \sum_{i=0_{\mathstrut}}^{N-1^{\mathstrut}}
S_{i}\tilde{S}_{i}^{\ast}=\mathbf{1}$\,, \\
for a dual \\
operator system \\
$\displaystyle S_{0},\dots ,S_{N-1}$, \\
$\displaystyle \smash{\tilde{S}_{0},\dots ,\tilde{S}_{N-1}}$} \\\hline
\end{tabular}
$

\addvspace{\bigskipamount}
\noindent
$\renewcommand{\arraystretch}{1.375}
\begin{tabular}
[c]
{p{0.40\textwidth}
|p{0.52\textwidth}}
\multicolumn{2}{c}{\parbox{0.9\textwidth}
{Then the assertions in the first table amount to:\rule[-12pt]{0pt}{12pt}}}\\
$\displaystyle C_{\psi}^{-1}\iint\limits_{\mathbb{R}^{2}}
\frac{dr\,ds}{r^{2}}
\left| \left\langle \,\psi_{r,s}\mid f\,\right\rangle \right| ^{2}
$\par
\hfill
$\displaystyle {}
=\left\Vert f\right\Vert _{L^{2}}^{2}\quad
\forall\,f\in L^{2}\left( \mathbb{R}\right) \vphantom{\iint}$
& $\displaystyle
C_{\!\psi,\tilde{\psi}}^{-1}\!\iint\limits_{\mathbb{R}^{2}}
\frac{dr\,ds}{r^{2}}
\left\langle \,f\mid \psi_{r,s}\,\right\rangle 
\left\langle \,\smash{\tilde{\psi}_{r,s}}\mid g\,\right\rangle \vphantom{\tilde{\psi}_{r,s}}
$\par
\hfill
$\displaystyle {}
=\left\langle \,f\mid g\,\right\rangle \quad
\forall\,f,g\in L^{2}\left( \mathbb{R}\right) \vphantom{\iint}$ \\\hline
$\displaystyle 
\sum_{\vphantom{k}j\in\mathbb{Z}}\sum_{\vphantom{j}k\in\mathbb{Z}_{\mathstrut}}
\left| \left\langle \,\psi_{j,k}\mid f\,\right\rangle \right| ^{2}
$\par
\hfill
$\displaystyle {}
=\left\Vert f\right\Vert_{L^{2}}^{2}\quad
\forall\,f\in L^{2}\left( \mathbb{R}\right) \vphantom{\iint}$
& $\displaystyle 
\sum_{\vphantom{k}j\in\mathbb{Z}}\sum_{\vphantom{j}k\in\mathbb{Z}_{\mathstrut}}
\left\langle \,f\mid \psi_{j,k}\,\right\rangle 
\left\langle \,\smash{\tilde{\psi}_{j,k}}\mid g\,\right\rangle \vphantom{\tilde{\psi}_{j,k}}
$\par
\hfill
$\displaystyle {}
=\left\langle \,f\mid g\,\right\rangle \quad
\forall\,f,g\in L^{2}\left( \mathbb{R}\right) \vphantom{\iint}$ \\\hline
$\displaystyle \sum_{i=0_{\mathstrut}}^{N-1^{\mathstrut}}
\left\Vert S_{i}^{\ast}c\right\Vert ^{2}=\left\Vert c\right\Vert ^{2}
\quad\forall\,c\in\ell^{2}$ 
& $\displaystyle \sum_{i=0_{\mathstrut}}^{N-1^{\mathstrut}}
\left\langle \,S_{i}^{\ast}c\mid \smash{\tilde{S}_{i}^{\ast}}d\,\right\rangle \vphantom{\tilde{S}_{i}^{\ast}}
=\left\langle \,c\mid d\,\right\rangle \quad\forall\,c,d\in\ell^{2}$ \\\hline
\end{tabular}
$

\addtolength{\textheight}{-0.5\baselineskip}%
\addvspace{\bigskipamount}

A
function $\psi$ satisfying the resolution identity is called
a \emph{coherent vector }
in mathematical physics. The
representation theory for the $\left( ax+b\right) $-group,
i.e., the matrix group
$G=\left\{\, \left( 
\begin{smallmatrix}
a & b \\
0 & 1
\end{smallmatrix}
\right) \mid a\in\mathbb{R}_{+},\; b\in\mathbb{R}\,\right\} $,
serves as its underpinning. Then the tables above
illustrate how the
$\left\{ \psi_{j,k}\right\} $ wavelet
system arises from a discretization of
the following unitary representation
of $G$:
\[
\left( U_{\left( 
\begin{smallmatrix}
a & b \\
0 & 1
\end{smallmatrix}
\right) }f\right) \left( x\right) 
=a^{-\frac{1}{2}}f\left( \frac{x-b}{a}\right) 
\]
acting on $L^{2}\left( \mathbb{R}\right) $.
This unitary representation
also explains

the discretization step in passing
from the first line to the second in
the tables above. The
functions $\left\{\, \psi_{j,k}\mid j,k\in\mathbb{Z}\,\right\} $ which
make up a wavelet system
result from the choice of a
suitable coherent
vector $\psi\in L^{2}\left( \mathbb{R}\right) $,
and then setting
\[
\psi_{j,k}\left( x\right) 
=\left( U_{\left( 
\begin{smallmatrix}
2^{-j} & k\cdot 2^{-j} \\
0 & 1
\end{smallmatrix}
\right) }\psi\right) \left( x\right) 
=2^{\frac{j}{2}}\psi\left( 2^{j}x-k\right) .
\]
Even though
this representation lies at the historical origin of the
subject of wavelets,
the $\left( ax+b\right) $-group seems
to be now largely forgotten
in the next generation of the wavelet community.
But Chapters 1--3 of \cite{Dau92}
still
serve as a beautiful presentation
of this (now much ignored) side
of the subject. It also serves
as a link to mathematical physics
and to classical analysis
.

\section{List of names and discoveries}
\label{sec3:List}
Many of the main
discoveries summarized below are now lore.

\begin{center}
\renewcommand{\arraystretch}{2}
\begin{tabular}{p{0.31\textwidth}@{\hspace*{0.04\textwidth}}p{0.62\textwidth}@{\hspace*{0.01\textwidth}}p{0.01\textwidth}}
{\raggedright\textbf{%
1807}\\Jean Baptiste Joseph Fourier
\\mathematics, physics
\\(heat conduction)}
&
Expressing functions as sums of sine and cosine waves of 
frequencies
in arithmetic progession (now called Fourier series).
\end{tabular}
\end{center}


\begin{center}
\renewcommand{\arraystretch}{2}
\begin{tabular}{p{0.31\textwidth}@{\hspace*{0.04\textwidth}}p{0.62\textwidth}@{\hspace*{0.01\textwidth}}p{0.01\textwidth}}
{\raggedright\textbf{%
1909}\\Alfred Haar
\\mathematics}  
& 
Discovered, while a student of 
David
Hilbert,
an orthonormal basis
consisting of step functions,
applicable 
both
to functions on an interval, and functions on the whole real line. While 
it
was not realized at the time, Haar's
construction
was a precursor of 
what is
now known as the Mallat subdivision,
and multiresolution
method, as well 
as
the subdivision wavelet
algorithms.
\\
{\raggedright\textbf{%
1946}\\Denes Gabor
\\(Nobel Prize): physics
\\(optics, holography)}
&
Discovered basis
expansions
for what might now be called time-frequency
wavelets,
as opposed to time-scale wavelets.
\\
{\raggedright\textbf{%
1948}\\Claude Elwood Shannon
\\mathematics, engineering
\\(information theory)}
& 
A rigorous formula used by the phone company for sampling
speech signals.
Quantizing
information, entropy, founder of what is now
called the mathematical theory of \mbox{communication.}
\\
{\raggedright\textbf{%
1976}\\Claude Garland, Daniel Esteban (both)\\signal processing}
&
Discovered subband
coding of digital transmission of speech signals
over 
the
telephone.
\\
{\raggedright\textbf{%
1981}\\Jean Morlet
\\petroleum engineer}
&
Suggested the term
``ondelettes.'' J.M. decomposed reflected seismic signals into sums of
``wavelets
(Fr.: ondelettes) of constant shape,'' i.e., a 
decomposition
of
signals into wavelet
shapes, selected from a library of such shapes (now
called wavelet
series). Received somewhat late recognition for his work. 
Due
to contributions by A. Grossman
and Y. Meyer,
Morlet's
discoveries 
have now
come to play a central role in the theory.
\\
{\raggedright\textbf{%
1985}\\Yves Meyer
\\mathematics,\\applications}
&
Mentor for A. Cohen,
S. Mallat,
and other of the wavelet
pioneers, Y.M.
discovered infinitely often differentiable
wavelets.
\\
{\raggedright\textbf{%
1989}\\Albert Cohen
\\mathematics (ortho-\\gonality relations),\\numerical
analysis}
&
Discovered the use of wavelet filters
in the 
analysis
of wavelets---the so-called Cohen
condition for orthogonality.
\end{tabular}
\end{center}

\begin{center}
\renewcommand{\arraystretch}{2}
\begin{tabular}{p{0.31\textwidth}@{\hspace*{0.04\textwidth}}p{0.62\textwidth}@{\hspace*{0.01\textwidth}}p{0.01\textwidth}}
{\raggedright\textbf{%
1986}\\St\'ephane Mallat
\\mathematics, signal\\and image processing}
&
Discovered 
what
is now known as the subdivision,
and multiresolution
method, as well as 
the
subdivision wavelet algorithms.
This allowed the effective use of 
operators
in the Hilbert space
$L^{2}(\mathbb{R})$\label{SLtwoRaw}, and of the parallel computational use of
recursive matrix algorithms.
\\
{\raggedright\textbf{%
1987}\\Ingrid Daubechies
\\mathematics, physics,
\\and communications}
&
Discovered differentiable
wavelets,
with the number of derivatives 
roughly
half the length of the support interval. Further found polynomial
algorithmic
 for their construction
(with coauthor Jeff Lagarias; joint
spectral radius formulas).
\\
{\raggedright\textbf{%
1991}\\Wayne Lawton
\\mathematics\\(the wavelet\\transfer
operator)}
&
Discovered the use of a transfer \mbox{operator}
in the 
analysis
of
wavelets: orthogonality
and smoothness.

\\
{\raggedright\textbf{%
1992}\\The FBI\\using wavelet algo-\\rithms
in digitizing and compressing
\\fingerprints}
&
C. Brislawn
and his group at Los Alamos 
created
the theory and the codes which allowed the compression of the enormous 
FBI
fingerprint file, creating A/D, a new database of fingerprints.
\\
{\raggedright\textbf{%
2000}\\The International\\Standards\\Organization}
&
A wavelet-based
picture compression standard, called JPEG 2000,
for \mbox{digital} encoding of
images.
\\
{\raggedright\textbf{%
1994}\\David Donoho\index{Donoho0@D. Donoho}\\statistics,
\\mathematics}
&
Pioneered the use 
of
wavelet bases
and tools from statistics
to ``denoise'' images
and 
signals.

\end{tabular}\label{endNames}
\end{center}

\section{History}
\label{sec4:Hist}

While wavelets as they have appeared in the mathematics literature (e.g., 
\cite{Dau92}) for a long time, starting with Haar in 1909, involve function 
spaces, the connections to a host of discrete problems from engineering is 
more subtle. Moreover the deeper connections between the discrete algorithms 
and the function spaces of mathematical analysis are of a more recent vintage, 
see e.g., \cite{StNg96} and \cite{Jor06a}.

Here we begin with the function spaces. This part of wavelet theory refers to 
continous wavelet transforms (details below). It dominated the wavelet 
literature in the 1980s, and is beautifully treated in the first four chapters 
in \cite{Dau92} and in \cite{Dau93}. The word ``continuous" refers to the 
continuum of the real line $\br$. Here we consider spaces of functions in one 
or more real dimensions, i.e., functions on the line $\br$ (signals), the 
plane $\br^{2}$ (images), or in higher dimensions $\br^{d}$, functions of $d$ 
real variables.

\section{Tools from Mathematics}
\label{sec6:Tools}
In our presentation, we will rely on tools from at least three separate areas 
of mathematics, and we will outline how they interact to form a coherent 
theory, and how they come together to form a link between what is now called 
the discrete and the continuous wavelet transform. It is the discrete case 
that is popular with engineers (\cite{AuKo06, Liu06, Str97, Str00}), while the 
continuous case has come to play a central role in the part of mathematics 
referred to as harmonic analysis, \cite{Dau93}. The three areas are, operator 
algebras, dynamical systems, and basis constructions:

\begin{itemize}
\item[a.]Operator algebras. The theory of operator algebras in turn breaks up 
in two parts: One the study of ``the algebras themselves" as they emerge from 
the axioms of von Neumann (von Neumann algebras), and Gelfand, Kadison and 
Segal ($C^{*}$-algebras.) The other has a more applied slant:
It involves ``the representations" of the algebras. By this we refer to the 
following: The algebras will typically be specified by generators and by 
relations, and by a certain norm-completion, in any case by a system of 
axioms. This holds both for the norm-closed algebras, the so called 
$C^{*}$-algebras, and for the weakly closed algebras, the von Neumann 
algebras. In fact there is a close connection between the two parts of the 
theory: For example, representations of $C^{*}$-algebras generate von Neumann 
algebras.

To talk about representations of a fixed algebra say $A$ we must specify a 
Hilbert space, and a homomorphism $\rho$ from $A$ into the algebra 
$\mathcal{B}(H)$ of all bounded operators on $\mathcal{H}$. We require that 
$\rho$ sends the identity element in $A$ into the identity operator acting on 
$\mathcal{H}$, and that $\rho(a^{*}) = (\rho(a))^{*}$ where the last star now 
refers to the adjoint operator.

It was realized in the last ten years (see for example \cite{BrJo02, Jor06a, Jor06b} 
that a family of representations that wavelets which are basis constructions 
in harmonic analysis, in signal/image analysis, and in computational 
mathematics may be built up from representations of an especially important 
family of simple $C^{*}$-algebras, the Cuntz algebras. The Cuntz algebras are 
denoted $\mathcal{O}_{2}, \mathcal{O}_3,...,$ including $\mathcal{O}_{\infty}$.

\item[b.] Dynamical systems. The connection between the Cuntz algebras 
$\mathcal{O}_{N}$ for $N= 2, 3,¡¦$ are relevant to the kind of dynamical 
systems which are built on branching-laws, the case of $\mathcal{O}_{N}$ 
representing $N$-fold branching. The reason for this is that if $N$ is fixed, 
$\mathcal{O}_{N}$ includes in its definition an iterated subdivision, but 
within the context of Hilbert space. For more details, 
see e.g., \cite{Dut04, DuRo07b, DuJo05, DuJo06a, DuJo06b, DuJo06c, Jor06b}.

\item[c.]  Analysis of bases in function spaces. The connection to basis 
constructions using wavelets is this: The context for wavelets is a Hilbert 
space $\mathcal{H}$, where $\mathcal{H}$ may be $L^{2}(\br^{d})$ where $d$ is 
a dimension, $d=1$ for the line (signals), $d=2$ for the plane (images), etc. 
The more successful bases in Hilbert space are the orthonormal bases ONBs, but 
until the mid 1980s, there were no ONBs in $L^{2}(\br^{d})$ which were 
entirely algorithmic and effective for computations. One reason for this is 
that the tools that had been used for $200$ years since Fourier involved basis 
functions (Fourier wave functions) which ere not localized. Moreover these 
existing Fourier tools were not friendly to algorithmic computations.

\end{itemize}

\section{A Transfer Operator}
\label{sec7:TO}
A popular tool for deciding if a candidate for a wavelet basis is in fact an 
ONB uses a certain transfer operator. Variants of this operator is used in 
diverse areas of applied mathematics. It is an operator which involves a 
weighted average over a finite set of possibilities. Hence it is natural for 
understanding random walk algorithms. As remarked in for example 
\cite{Jor03, Jor06a, Jor06b, Dut04}, it  was also studied in physics, for example by 
David Ruelle who used to prove results on phase transition for infinite spin 
systems in quantum statistical mechanics. In fact the transfer operator has 
many incarnations (many of them known as Ruelle operators), and all of them 
based on $N$-fold branching laws. 

In our wavelet application, the Ruelle operator weights in input over the $N$ 
branch possibilities, and the weighting is assigned by a chosen scalar 
function $W$. the  and the $W$-Ruelle operator is denoted $R_{W}$. In the 
wavelet setting there is in addition a low-pass filter function $m_{0}$ which 
in its frequency response formulation is a function on the $d$-torus 
$\mathbf{T}^{d} = \br^{d}/\bz^{d}$. 

Since the scaling matrix $A$ has integer entries $A$ passes to the quotient 
$\br^{d}/\bz^{d}$, and the induced transformation 
$r_{A} : \mathbb{T}^{d} \to \mathbb{T}^{d}$ is an $N$-fold cover, where 
$N = | det A|$, i.e., for every $x$ in $\mathbb{T}^{d}$ there are $N$ distinct 
points $y$ in $\mathbb{T}^{d}$ solving $r_{A}(y) = x$. 

In the wavelet case, the weight function $W$ is $W = |m_{0}|^{2}$. Then with 
this choice of $W$, the ONB problem for a candidate for a wavelet basis in the 
Hilbert space $L^{2}(\br^{d})$ as it turns out may be decided by the dimension 
of a distinguished eigenspace for $R_{W}$, by the so called Perron-Frobenius 
problem. 

This has worked well for years for the wavelets which have an especially 
simple algorithm, the wavelets that are initialized by a single function, 
called the scaling function. These are called the multiresolution analysis 
(MRA) wavelets, or for short the MRA-wavelets. But there are instances, for 
example if a problem must be localized in frequency domain, when the 
MRA-wavelets do not suffice, where it will by necessity include more than one 
scaling function. And we are then back to trying to decide if the output from 
the discrete algorithm, and the $\mathcal{O}_{N}$ representation is an ONB, or 
if it has some stability property which will serve the same purpose, in case 
where asking for an ONB is not feasible.

\section{Future Directions}
\label{sec8:Newdev}
The idea of a scientific analysis by subdividing a fixed picture or object into
its finer parts is not unique to wavelets.  It works best for structures with
an inherent self-similarity; this self-similarity can arise from numerical 
scaling of distances.  But there are more subtle non-linear self-similarities. 
The Julia sets in the complex plane are a case in point  
\cite{BY06, Br06, DeLo06, DeRo07, Mil04, PZ04}.  The simplest Julia set come
from a one parameter family of quadratic polynomials $\varphi_{c}(z)=z^{2}+c$,
where $z$ is a complex variable and where $c$ is a fixed parameter.  The 
corresponding Julia sets $J_{c}$ have a surprisingly rich structure.  A simple
way to understand them is the following: Consider the two brances of the 
inverse $\beta_{\pm}= z \mapsto \pm \sqrt{z-c}$.  Then $J_{c}$ is the unique
minimal non-empty compact subset of $\mathbb{C}$, which is invariant under 
$\{\beta_{\pm}\}$. (There are alternative ways of presenting $J_{c}$ but this
one fits our purpose. The Julia set $J$ of a holomorphic function, in this 
case $z \mapsto z^{2} + c$, informally consists of those points whose long-time 
behavior under repeated iteration , or rather iteration of substitutions, can 
change drastically under arbitrarily small perturbations.)  Here ``long-time" 
refers to largen $n$, where $\varphi^{(n+1)}(z)=\varphi(\varphi^{(n)}(z))$, 
$n=0,1,...$, and $\varphi^{(0)}(z)=z$.



It would be interesting to adapt and modify the Haar wavelet, and the other 
wavelet algorithms to the Julia sets.  The two papers \cite{DuJo05, DuJo06a} 
initiate such a development.

\section{Literature}
As evidenced by a simple Google check, the mathematical wavelet literature is
gigantic in size, and the manifold applications spread over a vast number of 
engineering journals.  While we cannot do justice to this volumest literature,
we instead offer a collection of the classics \cite{Hei06} edited recently by
C. Heil et.al.   

\begin{acknowledgements} 
We thank Professors Dorin Dutkay and Judy Packer for helpful discussions.
\end{acknowledgements}

\section{Bibliography}
\bibliographystyle{plain}
\bibliography{ency}

\begin{thebibliography}{10}

\bibitem{AuKo06}
Gilles Aubert and Pierre Kornprobst.
\newblock {\em Mathematical problems in image processing}.
\newblock 2006.

\bibitem{BJMP05}
Lawrence Baggett, Palle Jorgensen, Kathy Merrill, and Judith Packer.
\newblock A non-{MRA} {$C\sp r$} frame wavelet with rapid decay.
\newblock {\em Acta Appl. Math.}, 2005.

\bibitem{BrJo02}
0.~Bratelli and P.~Jorgensen.
\newblock {\em Wavelets Through a Looking Glass: The World of the Spectrum}.
\newblock Birkh\"{a}user, 2002.

\bibitem{BY06}
M.~Braverman and M.~Yampolsky.
\newblock Non-computable {J}ulia sets.
\newblock {\em J. Amer. Math. Soc.}, 19(3):551--578 (electronic), 2006.

\bibitem{Br06}
Mark Braverman.
\newblock Parabolic {J}ulia sets are polynomial time computable.
\newblock {\em Nonlinearity}, 19(6):1383--1401, 2006.

\bibitem{BrMa06}
K.~Bredies, D.~A. Lorenz, and P.~Maass.
\newblock {\em An optimal control problem in medical image processing}.
\newblock 2006.

\bibitem{Dau92}
Ingrid Daubechies.
\newblock {\em Ten lectures on wavelets}, volume~61 of {\em CBMS-NSF Regional
  Conference Series in Applied Mathematics}.
\newblock 1992.

\bibitem{Dau93}
Ingrid Daubechies.
\newblock {\em Wavelet transforms and orthonormal wavelet bases}.
\newblock Proc. Sympos. Appl. Math. 1993.

\bibitem{DaLa92}
Ingrid Daubechies and Jeffrey~C. Lagarias.
\newblock Two-scale difference equations. {II}. {L}ocal regularity, infinite
  products of matrices and fractals.
\newblock {\em SIAM J. Math. Anal.}, 1992.

\bibitem{DeLo06}
Robert~L. Devaney and Daniel~M. Look.
\newblock A criterion for {S}ierpinski curve {J}ulia sets.
\newblock {\em Topology Proc.}, 30(1):163--179, 2006.
\newblock Spring Topology and Dynamical Systems Conference.

\bibitem{DeRo07}
Robert~L. Devaney, M{\'o}nica~Moreno Rocha, and Stefan Siegmund.
\newblock Rational maps with generalized {S}ierpinski gasket {J}ulia sets.
\newblock {\em Topology Appl.}, 154(1):11--27, 2007.

\bibitem{DuJo06c}
Dorin~E. Dutkay and Palle E.~T. Jorgensen.
\newblock Wavelets on fractals.
\newblock {\em Rev. Mat. Iberoamericana}, 22, 2006.

\bibitem{DuRo07b}
Dorin~E. Dutkay and Kjetil Roysland.
\newblock The algebra of harmonic functions for a matrix-valued transfer
  operator.
\newblock {\em arXiv:math/0611539}, 2007.

\bibitem{DuRo07a}
Dorin~E. Dutkay and Kjetil Roysland.
\newblock Covariant representations for matrix-valued transfer operators.
\newblock {\em arXiv:math/0701453}, 2007.

\bibitem{Dut04}
Dorin~Ervin Dutkay.
\newblock The spectrum of the wavelet {G}alerkin operator.
\newblock {\em Integral Equations Operator Theory}, 2004.

\bibitem{DuJo05}
Dorin~Ervin Dutkay and Palle E.~T. Jorgensen.
\newblock Wavelet constructions in non-linear dynamics.
\newblock {\em Electron. Res. Announc. Amer. Math. Soc.}, 2005.

\bibitem{DuJo06a}
Dorin~Ervin Dutkay and Palle E.~T. Jorgensen.
\newblock Hilbert spaces built on a similarity and on dynamical
  renormalization.
\newblock {\em J. Math. Phys.}, 2006.

\bibitem{DuJo06b}
Dorin~Ervin Dutkay and Palle E.~T. Jorgensen.
\newblock Iterated function systems, {R}uelle operators, and invariant
  projective measures.
\newblock {\em Math. Comp.}, 2006.

\bibitem{Hei06}
Christopher Heil and David~F. Walnut, editors.
\newblock {\em Fundamental papers in wavelet theory}.
\newblock Princeton University Press, Princeton, NJ, 2006.

\bibitem{Jor03}
Palle E.~T. Jorgensen.
\newblock Matrix factorizations, algorithms, wavelets.
\newblock {\em Notices Amer. Math. Soc.}, 50, 2003.

\bibitem{Jor06a}
Palle E.~T. Jorgensen.
\newblock {\em Analysis and probability: wavelets, signals, fractals}, volume
  234 of {\em Graduate Texts in Mathematics}.
\newblock Springer, New York, 2006.

\bibitem{Jor06b}
Palle E.~T. Jorgensen.
\newblock Certain representations of the {C}untz relations, and a question on
  wavelets decompositions.
\newblock 414:165--188, 2006.

\bibitem{Liu06}
Feng Liu.
\newblock Diffusion filtering in image processing based on wavelet transform.
\newblock {\em Sci. China Ser. F}, 49, 2006.

\bibitem{Mil04}
John Milnor.
\newblock Pasting together {J}ulia sets: a worked out example of mating.
\newblock {\em Experiment. Math.}, 13(1):55--92, 2004.

\bibitem{PZ04}
C.~L. Petersen and S.~Zakeri.
\newblock On the {J}ulia set of a typical quadratic polynomial with a {S}iegel
  disk.
\newblock {\em Ann. of Math. (2)}, 159(1):1--52, 2004.

\bibitem{Sko01}
A.~Skodras, C.~Christopoulos, and T.~Ebrahimi.
\newblock Jpeg 2000 still image compression standard" ieee signal processing
  magazine.
\newblock {\em IEEE Signal processing Magazine}, 18:36--58, Sept. 2001.

\bibitem{Son05}
Myung-Sin Song.
\newblock {\em Wavelet image compression}.
\newblock Ph.D. thesis. The University of Iowa, 2006.

\bibitem{Son06}
Myung-Sin Song.
\newblock Wavelet image compression.
\newblock In {\em Operator theory, operator algebras, and applications}, volume
  414 of {\em Contemp. Math.}, pages 41--73. Amer. Math. Soc., Providence, RI,
  2006.

\bibitem{Str97}
Gilbert Strang.
\newblock {\em Wavelets from filter banks}.
\newblock Springer, 1997.

\bibitem{Str00}
Gilbert Strang.
\newblock {\em Signal processing for everyone}, volume 1739 of {\em Lecture
  Notes in Math.}
\newblock 2000.

\bibitem{StNg96}
Gilbert Strang and Truong Nguyen.
\newblock {\em Wavelets and filter banks}.
\newblock Wellesley-Cambridge Press, Wellesley, MA, 1996.

\bibitem{Use01}
B.~E. Usevitch.
\newblock A tutorial on modern lossy wavelet image compression: Foundations of
  jpeg 2000.
\newblock {\em IEEE Signal processing Magazine}, 18:22--35, Sept. 2001.

\bibitem{Wal99}
J.~S Walker.
\newblock {\em A Primer on Wavelets and their Scientific Applications}.
\newblock Chapman \& Hall, CRC, 1999.

\end{thebibliography}

\end{document}